\documentclass[conference]{IEEEtran}

\usepackage{fancyhdr}
\usepackage{hyperref}
\usepackage{epsfig}
\usepackage{algorithm}
\usepackage{algorithmic}
\usepackage{graphicx}
\usepackage{amsmath,amsfonts,amssymb}
\usepackage{color}
\usepackage{multirow}
\usepackage{epstopdf}
\usepackage{bbm}
\usepackage[table]{xcolor}
\usepackage[square, numbers]{natbib}
\usepackage{amsthm}
\usepackage{cancel}
\usepackage{subfig}

\newtheorem{thm}{Theorem}[section]
\newtheorem{lem}[thm]{Lemma}
\newtheorem{cor}[thm]{Corollary}
\newenvironment{pf}{\noindent{\bf Proof:} \hspace*{1mm}}{ \hspace*{\fill} $\Box$ }

\newcommand{\argmax}{\arg\,\max}

\begin{document}
\title{\vspace{-0.2in}Game Theoretic Analysis of Tree Based Referrals for Crowd Sensing Social Systems with Passive Rewards
\vspace{-0.2in}}
\author{\IEEEauthorblockN{Kundan Kandhway and Bhushan Kotnis}
\IEEEauthorblockA{Department of Electronic Systems Engineering, Indian Institute of Science\\
Email: \{kundan,bkotnis\}@dese.iisc.ernet.in}
}


\maketitle

\thispagestyle{fancy}

\begin{abstract}
Participatory crowd sensing social systems rely on the participation of large number of individuals. Since humans are strategic by nature, effective incentive mechanisms are needed to encourage participation. A popular mechanism to recruit individuals is through referrals and passive incentives such as geometric incentive mechanisms used by the winning team in the 2009 DARPA Network Challenge and in multi level marketing schemes. The effect of such recruitment schemes on the effort put in by recruited strategic individuals is not clear. This paper attempts to fill this gap. Given a referral tree and the direct and passive reward mechanism, we formulate a network game where agents compete for finishing crowd sensing tasks. We characterize the Nash equilibrium efforts put in by the agents and derive closed form expressions for the same.  We discover free riding behavior among nodes who obtain large passive rewards. This work has implications on designing effective recruitment mechanisms for crowd sourced tasks. For example, usage of geometric incentive mechanisms to recruit large number of individuals may not result in proportionate effort because of free riding. 

\end{abstract}

\section{Introduction} \label{sec:intro}
The widespread presence of smart phones, wearables, GPS devices and other hand held sensors have enabled individuals to collect valuable data from their surrounding environments.  Analyzing  such crowd sensed data can help monitor urban and industrial pollution levels \cite{Ganti2011}, improve our understanding of urban traffic patterns \cite{Pan2013}, and can even prove to be useful for surveillance and emergency response \cite{Chu2012}. Some crowd sensing tasks, such as traffic monitoring, can be performed using opportunistic sensing that does not actively involve the user, while others such as pollution monitoring or surveillance may require active participation of the user \cite{Dutta2009}. Such participatory sensing systems, due to the human in the loop, raises unique challenges such as recruiting and incentivizing a large number of individuals to participate in the sensing activity. 

Offering referral rewards for recruiting friends to sign up for completing a task is widely used in multi level marketing \cite{Emek2011}. Such mechanisms generate referral trees where a parent node represents an individual that has recruited or referred another individual who is represented as a child to the referring parent node. In such schemes, apart from rewarding individuals for finishing the tasks, additional rewards, in the form of incentives, are provided to encourage individuals to recruit people from their social connections. A popular strategy is to provide a proportion of the reward earned by recruited individuals to the recruiting individual recursively.  Such  mechanisms, where the parent node obtains a proportion of the reward earned by the child nodes, are termed as geometric incentive mechanisms \cite{Emek2011}  and such indirect rewards are called passive rewards. 

Geometric incentive mechanisms have also been used with success in participatory crowd sourcing tasks. The team that won the 2009 DARPA  Network  Challenge used a geometric incentive mechanism to recruit a large number of volunteers to complete the task \cite{Pickard2011}. The challenge involved locating the positions of 10 large red weather balloons scattered in the continental United States in the shortest possible time \cite{Pickard2011}. The prize winning team developed a mechanism where an individual who finds the location of the balloon passes half of her reward to her parent (recruiter), who in turn passes half of it to her parent, and so on.  Thus, an ancestor who is $k$ hop away from the node who finished the task gets $\frac{1}{2^{k+1}}$ part of the reward (the mechanism may restrict reward sharing to only a few levels). A mathematical analysis of the of this strategy revealed that for a rational individual recruiting the maximum number of people she knows is the best response strategy \cite{Pickard2010}. 

Clearly for such a large search task, a large number of recruits is key. However, that is not the only important factor; the efforts put in by those individuals to finish the task is equally critical. Particularly, the incentive mechanism which was used to recruit individuals should also ensure that it does not disincentivize individuals to work hard. The effect of geometric incentive mechanisms, that were used to recruit individuals, on the efforts put in by them to finish the task is not clear. In this paper we aim to fill that gap.

By formulating a network game we show the effect geometric incentive mechanisms have on the  efforts put in by rational agents to finish the task. We analyze the behavior of agents connected over a referral network (tree) and compute an equilibrium effort profile. Our analysis technique is game theoretic, i.e., we model each individual as a strategic agent who wants to maximize her utility in a given situation. The equilibrium we consider is a pure strategy Nash equilibrium (PSNE), and we show that it always exists and is unique for interesting model parameters. More importantly, our results show that while geometric incentive schemes are an excellent tool for recruiting individuals they may indeed disincentivize a few individuals from putting any effort.

\section{Model \label{sec:model}}
\vspace{-0.05in}
\subsection{The Task Arrival and Processing Model} 
Let the set of individuals be denoted by $N = \{1,2,\dots, n\}$, and connected over a hierarchical directed tree $T$. The direction of the edges are from the root towards the leaves. We assume that if an individual performs a crowd sourced task, she gets a direct incentive, and the nodes along the path from the individual to the root receive an indirect incentive. We assume that the tasks arrive at a Poisson rate $\lambda$, which is a common knowledge, and are queued up until served. We assume that each agent competes for the task (as is the case in multi level marketing and the 2009 DARPA Network Challenge). Each node $i$ attempts to capture a task from the work queue according to a Poisson process with rate $\lambda_i$. If the total reward of a task is $R$, we assume that a node gets a direct incentive of $\gamma \cdot R$ if she grabs the job ($\gamma \in (0,1)$), with the rest shared as indirect incentives. The cost of maintaining an attempt rate $\lambda_i$ is $C \cdot \lambda_i$, where $C >0$ is a known constant. The task is assigned exclusively to the agent who attempts it first after the task arrives. We assume the agents have uniform skill of performing a task. The time to complete a task is small compared to the inter-attempt times of any agent and hence the task completion time is ignored.

We can model the task arrival and departure in a server-queue model where the arrival rate is $\lambda ( > 0)$ and the consolidated service rate of the entire network is $\sum_{j \in N} \lambda_j$, because the superposition of Poisson processes is Poisson with the rate being the sum of the rates~\cite{Bertsekas1992}. When agent $i$ tries to capture a job from the task queue, the probability that she can grab the job is given by $\lambda_i / \sum_{j \in N} \lambda_j$. This is because the inter-attempt times are exponentially distributed for a Poisson process, and using the memoryless property of exponential random variables \cite{Wolff1989}.
There are two regimes the system could operate in.

\noindent {\bf Case 1}: When $\sum_{j \in N} \lambda_j > \lambda$, the queueing process is a positive recurrent Markov chain and all tasks will be served, and the output process would also be Poisson with rate $\lambda$ (Burke's Theorem \cite{Bertsekas1992}), the rate at which agent $i$ would grab the task is therefore given by $\lambda \cdot \lambda_i / \sum_{j \in N} \lambda_j$. Hence, the direct reward to $i$ is $\lambda \gamma R \cdot \lambda_i / \sum_{j \in N} \lambda_j$. 

\noindent {\bf Case 2}: When $\sum_{j \in N} \lambda_j \leq \lambda$, the queueing process is a null or transient Markov chain and with high probability (probability approaching unity) the queue will be non-empty at a steady state~\cite{Wolff1989}. In such a setting, any agent $i$ who attempts to grab a task actually gets a task, and hence the direct reward of agent $i$ would be $\gamma R \cdot \lambda_i$.

If a node $j$ hits a task, she receives a `direct' reward, and each node $i$ on the directed path from the root to $j$ receives an `indirect' reward. Since the efforts are costly ($C \lambda_i$), each agent has to decide on the efforts to maximize her net payoff, i.e., (direct + indirect) reward minus the cost. If there are indirect rewards, a node will reduce efforts to reduce costs. This induces a game between the nodes. The strategy of player $i$ is to choose the attempt rate $\lambda_i \in S_i = [0, \infty)$. In the following section, we derive an expression for the expected utility for an agent $i$.

\vspace{-.03in}
\subsection{Indirect Reward and the Utility Model} \label{sec:network-models}

\vspace{-.04in}
The task arrival process and its incentive sharing scheme induce competition among the nodes. Direct incentives are earned when an agent grabs a task. 

The indirect incentives are shared with other players as specified by the reward sharing matrix $\Delta = [\delta_{ij}]$ where $\delta_{ij}$ is the fraction of the total reward received by $i$ when $j$ grabs the task. Note that $\delta_{ii} = \gamma, \ \forall i \in N$. Since the network is a directed tree, if $j$ does not appear in subtree of $i$ then $\delta_{ij} = 0$. We call the matrix $\Delta$ {\em monotone non-increasing} if $\delta_{ij} > \delta_{ij'}$, whenever the hop-distance $\text{dist}_T(i,j) < \text{dist}_T(i,j')$, and {\em anonymous} if the $\delta_{ij}$ depends only on the distance of $i$ and $j$ on $T$, $\text{dist}_T(i,j)$, and not on the identities of the nodes. Hence these are reasonable assumptions to make. 

While some of our results extend to general networks and general $\Delta$ matrices, in this paper, we focus on monotone non-increasing and anonymous reward sharing matrices. To ensure that the total reward is bounded above by $R$, we assume $\sum_{k \in N} \delta_{kj} \leq 1, \ \forall j \in N$. Let us denote $T_i$ as the subtree rooted at $i$. Combining the direct and indirect rewards and costs, the expected utility of agent $i \in N$ given by, 
\begin{equation} \label{eqn:util-T}
 u_i(\lambda_i,\lambda_{-i}) = \left \{    
    \begin{array}{lr}
      u_{i,1}(\lambda_i,\lambda_{-i}), & \qquad \sum_{k \in N}\lambda_k > \lambda, \\
      u_{i,2}(\lambda_i,\lambda_{-i}), & \qquad \sum_{k \in N}\lambda_k \leq \lambda,
    \end{array}
\right.
\end{equation}
where $\lambda_{-i} = (\lambda_1,\dots,\lambda_{i-1},\lambda_{i+1},\dots,\lambda_n)$, and, 
\footnotesize
\begin{align}
 u_{i,1}(\lambda_i,\lambda_{-i}) &:= \lambda \gamma R \frac{\lambda_i}{\sum_{k \in N}\lambda_k} 
    + \lambda R \hspace{-1em} \sum_{j \in T_i \setminus \{i\}} \hspace{-1em}\delta_{ij}  \frac{\lambda_j}{\sum_{k \in N}\lambda_k} 
    - C \lambda_i, \label{eqn:util1-T} \\
 u_{i,2}(\lambda_i,\lambda_{-i}) &:= \gamma R \lambda_i 
    + \lambda R \hspace{-1em}\sum_{j \in T_i \setminus \{i\}} \hspace{-1em}\delta_{ij} \lambda_j
    - C \lambda_i. \label{eqn:util2-T}
\end{align}
\normalsize
In both (\ref{eqn:util1-T}) and (\ref{eqn:util2-T}), the first term on the RHS denotes the expected utility of agent $i$ due to her own effort $\lambda_i$, the second term denotes the indirect utility coming from the efforts $\lambda_j$ of all the nodes $j \in T_i \setminus \{i\}$, and the third term denotes the cost to maintain the effort. The utility model is parametrized by the reward $R$, reward sharing matrix $\Delta$, cost $C$, and tree $T$.

\subsection{Effort Sharing and Effort Zones}
\vspace{-.025in}

The {\em effort sharing function} $f$ is a recursive function, which is computed for a given tree $T$ and a reward sharing matrix $\Delta$, in a bottom up fashion, i.e., from the leaves towards the root. Let us denote the set of directed trees by ${\cal T}$. For any tree, $f$ is initialized to $1$ for all the leaves, and then computed as follows.

\noindent {\bf Effort Sharing Function}: A mapping $f : {\cal T} \times [0,1)_{n \times n} \to [0,1]$ given by the recursive formula,
\begin{align} \label{eqn:f-def}
 \mbox{$f(T_i,\Delta) = \max \left\{0, 1 - \sum_{j \in T_i \setminus \{i\}} \delta_{ij} \cdot f(T_j,\Delta) \right\}$.}
\end{align}
This function is maximum when $i$ is a leaf, and decreases for nodes with large subtrees below them (i.e., the ones with large passive rewards). In Sec. \ref{sec:Analysis} we will see that, for interesting model parameters, the effort equilibrium levels are proportional to this function, leading to smaller effort levels for nodes with large subtrees below them. Based on this we partition the space of parameters $V := \{\mathbf{v} = (R, \Delta, C, T)\}$ into four regions:
\vspace{-.06in}
\begin{align}
& \mbox{Region I: } & \mbox{$R_1$} &= \mbox{$\left \{\mathbf{v} : \frac{\gamma \cdot R}{C} \in [0, 1) \right \}$} \label{eqn:R1} \\
& \mbox{Region II: } & \mbox{$R_2$} &= \mbox{$\left \{\mathbf{v} : \frac{\gamma \cdot R}{C} = 1 \right \}$} \label{eqn:R2} \\
& \mbox{Region III: } &  \mbox{$R_3$} &= \mbox{$\left \{\mathbf{v} : \frac{\gamma \cdot R}{C} \in \left(1, \frac{\sum_{j \in N} f(T_j,\Delta)}{\sum_{j \in N} f(T_j,\Delta) - 1} \right] \right \}$} \label{eqn:R3} \\
& \mbox{Region IV: } & \mbox{$R_4$} &= \mbox{$\left \{\mathbf{v} : \frac{\gamma \cdot R}{C} > \frac{\sum_{j \in N} f(T_j,\Delta)}{\sum_{j \in N} f(T_j,\Delta) - 1}\right \}$} \label{eqn:R4}
\end{align}
We will analyze PSNE effort profiles for the game in the whole parameter space. We will see in Sec. \ref{sec:Analysis} that the game exhibits different behavior in each of these regions (and hence the regions are divided in this way).

\noindent {\bf Effort Zones}: We partition the space of agents' effort profile $L := \{\boldsymbol\lambda = (\lambda_1, \dots, \lambda_n)\}$ into four regions:
\begin{align}
\mbox{$Z_1$} &= \mbox{$\left \{ \boldsymbol\lambda : \sum_{j \in N} \lambda_j < \lambda \right \}$} \qquad \qquad \mbox{ Zone 1,} \label{eqn:Z1} \\
\mbox{$Z_2$} &= \mbox{$\left \{ \boldsymbol\lambda : \sum_{j \in N} \lambda_j = \lambda \right \}$} \qquad \qquad \mbox{ Zone 2,} \label{eqn:Z2} \\
\mbox{$Z_3$} &= \mbox{$\Big\{\boldsymbol \lambda \in Z_2: \lambda_i + \sum_{j \in T_i \setminus \{i\}} \delta_{ij} \lambda_j $} \nonumber \\
& \mbox{ $ \geq \lambda \left( 1 - \frac{C}{\gamma R} \right), \forall \ i \in N\Big\}$}  \qquad \ \ \ \mbox{ Zone 3,} \label{eqn:def-Z_3}\\
\mbox{$Z_4$} &= \mbox{$\left \{ \boldsymbol\lambda : \sum_{j \in N} \lambda_j > \lambda \right \}$} \qquad \qquad \mbox{ Zone 4.} \label{eqn:Z4}
\end{align}
Notice that $Z_3 \subseteq Z_2$. We define it separately to show a result that characterizes the set of Nash equilibria for region $R_3$. If we consider a single server abstraction of the entire network, then the arrival rate $\lambda$ sees a consolidated service rate of $\sum_{i \in N} \lambda_i$. If the service rate is smaller, equal, or larger than the arrival rate, then according to the conditions (\ref{eqn:Z1}--\ref{eqn:Z4}), the task queue would be either over loaded (Zone 1), critically loaded (Zone 2, Zone 3), or under loaded (Zone 4) respectively.

The crowd sourcing manager would like to operate in $Z_4$, so that over a long period of time, \emph{all} incoming tasks are served and there is no accumulation of tasks.

We see from (\ref{eqn:util-T}) that the zones correspond to different utility structures, \\
(i) $u_i(\boldsymbol\lambda) = u_{i,1}^{\text{T}}(\boldsymbol\lambda)$, if $\boldsymbol\lambda \in Z_4$; \ \ \ \ \ \ \ \ (ii) $u_i(\boldsymbol\lambda) = u_{i,2}^{\text{T}}(\boldsymbol\lambda)$, if $\boldsymbol\lambda \in Z_1 \cup Z_2$.

\begin{figure}
\begin{center}
\includegraphics[width=0.35\textwidth]{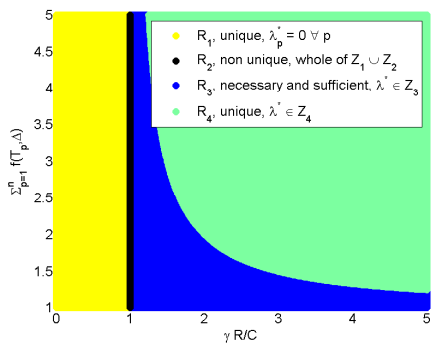}
\end{center}
\caption{\small Graphical illustration of the parameter space and the Nash equilibria characterization.}
\label{fig:param_space}
\vspace{-0.25in}
\end{figure}

In the following section, we show that in each of the regions $R_1$ to $R_4$, the PSNE are mapped to the zones $Z_1$ to  $Z_4$, sometimes uniquely. The PSNE always exists, which is a non-trivial result. Fig.~\ref{fig:param_space} graphically illustrates the parameter space and the main results on the characterization of the equilibria (detailed analysis is carried out in Sec. \ref{sec:Analysis}). It gives the intuition that since the reward to cost ratio $R/C$ is small in regions $R_1$ to $R_3$, the sum of the equilibrium efforts could fall below the incoming rate $\lambda$. However, in region $R_4$, the $R/C$ ratio is large enough to ensure a unique PSNE, in which the sum equilibrium effort can efficiently serve the incoming task stream.

\section{Analytical Results} \label{sec:Analysis}
We compute the PSNE for the four regions $R_1$ to $R_4$ in Eqn (\ref{eqn:R1})--(\ref{eqn:R4}) (hence a PSNE always exists). The results serve to predict the efforts put in by strategic agents connected via a network for the given reward sharing matrix $\Delta$.

The first result shows that $u_i$ is concave in $\lambda_i$. This helps in calculation of the PSNE as it is a utility maximization process for all agents.

\begin{lem} \label{claim:concave-ui}
$u_i(\lambda_i, \lambda_{-i})$ is a concave function in $\lambda_i$.
\end{lem}
\begin{pf}
We show this in two steps.

\textbf{\emph{Step 1}}: $u_{i,2}(\lambda_i, \lambda_{-i})$ is a linear function in $\lambda_i$, which is a special case of a concave function. Now we show that $u_{i,1}(\lambda_i, \lambda_{-i})$ is concave in $\lambda_i$. Differentiating Eq. (\ref{eqn:util1-T}),
\begin{equation}
\mbox{$\frac{\partial}{\partial \lambda_i}u_{i,1}(\lambda_i, \lambda_{-i}) = \frac{\lambda \gamma R \left(\sum_{j \neq i}\lambda_j - \sum_{j \in T_i \setminus \{i\}} \delta_{ij} \lambda_j \right)}{{\left(\sum_{j \in N} \lambda_j \right)}^2} - C$}
\label{du1bydlami}
\end{equation}
and by taking the second derivative we get,
\begin{equation}
\mbox{$\frac{\partial^2}{\partial \lambda_i^2}u_{i,1}(\lambda_i, \lambda_{-i}) = \frac{-2\lambda \gamma R \left(\sum_{j\neq i} \lambda_j - \sum_{j \in T_i \setminus \{i\}} \delta_{ij} \lambda_j \right)}{{\left(\sum_{j \in N} \lambda_j \right)}^3}$}
\end{equation}
Since $\lambda_i \geq 0 \ \forall i, \left(\sum_{j\neq i} \lambda_j - \sum_{j \in T_i \setminus \{i\}} \delta_{ij} \lambda_j \right) \geq 0$, this is due to the fact that $\delta_{ij} \in [0,1)$ and $|\{j \in N : j \neq i \}| \geq |T_i \setminus \{i\}|$. Hence, $u_{i,1}(\lambda_i, \lambda_{-i})$ is concave in $\lambda_i$.

\textbf{\emph{Step 2}}: We observe that, $u_i(\lambda_i, \lambda_{-i}) = \textnormal{min}\{u_{i,1}(\lambda_i, \lambda_{-i}), u_{i,2}(\lambda_i, \lambda_{-i})\}$.
This is because that when $\frac{\lambda}{\sum_{j \in N} \lambda_j} < 1$, $~~u_{i,1}(.) < u_{i,2}(.)$ and when $\frac{\lambda}{\sum_{j \in N}\lambda_j} \geq 1$, $u_{i,1}(.) \geq u_{i,2}(.)$  from Equations (\ref{eqn:util-T}), (\ref{eqn:util1-T}) and (\ref{eqn:util2-T}). The function $u_i(\cdot)$ is the minimum of two concave functions and hence is concave.
\end{pf}

\noindent The next result characterizes the PSNE in region $R_1$.

\begin{thm}[PSNE in $R_1$]
\label{theorem:equilibrium-in-R1}
If $\mathbf{v} \in R_1$, then the PSNE effort profile is unique, and is given by $\boldsymbol\lambda^* = (0,\dots,0)$.
\end{thm}
\begin{pf}
When $\mathbf{v} \in R_1$, we argue that all players will put zero effort in Nash Equilibrium. Suppose $\boldsymbol \lambda^* \in Z_4$, it implies $u_i(\boldsymbol \lambda^*) = u_{i,1}(\boldsymbol \lambda^*)$. Taking derivative of the utility at $\boldsymbol \lambda^*$, we get,

\vspace{-.2in}
\small
\begin{align*}
\lefteqn{\frac{\partial}{\partial \lambda^*_i}u_{i,1}(\lambda^*_i, \lambda^*_{-i})
= \frac{\lambda \gamma R \left(\sum_{j \neq i}\lambda^*_j - \sum_{j \in T_i \setminus \{i\}} \delta_{ij} \lambda^*_j \right)}{{\left(\sum_{j \in N} \lambda^*_j \right)}^2} - C} \\
&= \underbrace{\left(\frac{\lambda}{\sum_{j \in N} \lambda^*_j} \right)}_{< 1} \cdot \gamma R \cdot \underbrace{\frac{\left(\sum_{j \neq i}\lambda^*_j - \sum_{j \in T_i \setminus \{i\}} \delta_{ij} \lambda^*_j \right)}{\sum_{j \in N} \lambda^*_j}}_{\leq 1} -C \\
&< \gamma R - C < 0.
\end{align*}
\normalsize

\noindent This is a contradiction for $\boldsymbol \lambda^*$ to be an equilibrium, since each player $i$ would be better off by decreasing their effort from $\lambda_i^*$. Hence $\boldsymbol \lambda^* \notin Z_4$. So, $\boldsymbol \lambda^* \in Z_1 \cup Z_2$, which implies $u_i(\boldsymbol \lambda^*) = u_{i,2}(\boldsymbol \lambda^*)$. But $\frac{\partial}{\partial\lambda_i}u_{i,2} = \gamma R - C < 0$, since $\mathbf{v} \in R_1$. Thus $\lambda_i^* = 0$ for all $i \in N$.
\end{pf}

The intuition is that in $R_1$, the reward to cost ration $R/C$ is small enough. Hence, no individual gets any positive payoff by putting any positive effort. In $R_2$, the $R/C$ ratio reaches a critical value, where there exists multiple PSNE. The agents collectively are indifferent between just meeting the incoming rate $\lambda$ and keeping the sum effort smaller than this. Hence we get the following theorem.

\begin{thm}[PSNE in $R_2$]
\label{theorem:equilibrium-in-R2}
If $\mathbf{v} \in R_2$, then any $\boldsymbol \lambda^* \in Z_1 \cup Z_2$ is a PSNE. 
\end{thm}
\begin{pf}
\textbf{\emph{Case 1}}: Suppose $\boldsymbol \lambda^* \in Z_4$. Then $u_i(\boldsymbol \lambda^*) = u_{i,1}(\boldsymbol \lambda^*)$ and we repeat the analysis of Theorem~\ref{theorem:equilibrium-in-R1} to show that $\frac{\partial}{\partial \lambda^*_i}u_{i,1}(\lambda^*_i, \lambda^*_{-i}) < 0, \forall \ i \in N$, even for $\mathbf{v} \in R_2$. Hence it is not an equilibrium as players will keep decreasing their efforts.
\textbf{\emph{Case 2}}: When $\boldsymbol \lambda^* \in Z_1 \cup Z_2$, we have $u_i(\boldsymbol \lambda^*) = u_{i,2}(\boldsymbol \lambda^*)$ and $\frac{\partial}{\partial \lambda^*_i}u_{i,2}(\lambda^*_i, \lambda^*_{-i}) = 0, \forall \ i \in N$. Thus, $u_i(\lambda_i, \lambda_{-i}^*)$ is constant when $\sum_{j \in N} \lambda_j^* \leq \lambda$, for all $i \in N$. Hence each element of the set $Z_1 \cup Z_2$ is a PSNE.
\end{pf}

For the sake of convenience, \textbf{we will discuss region $R_4$, before $R_3$.} In the final region $R_4$, the ratio $R/C$ crosses a minimum threshold, which guarantees that the equilibrium effort profile is unique and sufficient to serve the incoming task rate, i.e., $\sum_{j \in N} \lambda_j^* > \lambda$. We show this using a few intermediate lemmas.

\begin{lem} \label{lemma:R4-notin}
If $\mathbf{v} \in R_4$, and if there exists a PSNE effort profile $\boldsymbol\lambda^*$, it cannot lie in $Z_1 \cup Z_2$.
\end{lem}
\begin{pf}
We prove this via contradiction. Suppose $\exists$ a PSNE effort profile $\boldsymbol\lambda^* \in Z_1 \cup Z_2$, i.e., $\sum_{j \in N} \lambda_j^* \leq \lambda$. Hence $u_i(\lambda_i^*, \lambda_{-i}^*) = u_{i,2}(\lambda_i^*, \lambda_{-i}^*), \forall \ i \in N$. But, $\frac{\partial}{\partial\lambda_i}u_{i,2}(\boldsymbol \lambda^*) = \gamma R - C > 0, \forall \ i \in N$ (differentiating Eq. (\ref{eqn:util2-T})). This is because, $\mathbf{v} \in R_4$ and from Eqn (\ref{eqn:R4}) we know that $\frac{\gamma R}{C} > \frac{\sum_{j \in N} f(T_j,\Delta)}{\sum_{j \in N} f(T_j,\Delta) - 1} > 1$. Hence, utility of $i$ is increasing in $\lambda_i$ at $\boldsymbol \lambda^*$. So, $i$ is better off by increasing his effort from $\lambda^*_i$, which is a contradiction to the fact that $\lambda^*$ is a PSNE.
\end{pf}
\begin{cor}
\label{corollary:util_in_R4}
 For $\mathbf{v} \in R_4$, if there exists a PSNE effort profile $\boldsymbol \lambda^*$ then $u_i(\boldsymbol \lambda^*) = u_{i,1}(\boldsymbol\lambda^*), \ \forall \ i \in N$.
\end{cor}

\begin{lem} \label{lemma:R4-exact-expr}
 If the utility structure $u_i$ is given by $u_{i,1}, \ \forall \ i \in N$, then there exists a unique PSNE effort profile $\boldsymbol\lambda^*$ given by,
 \begin{align} \label{eqn:R4-lambda}
  \mbox{$\lambda_i^* = \frac{\lambda \gamma R}{C}\left( \frac{\sum_{j \in N} f(T_j,\Delta) - 1}{\left(\sum_{j \in N} f(T_j,\Delta)\right)^2} \right) f(T_i,\Delta), \forall \ i \in N$},
 \end{align}
where the function $f$ is defined in (\ref{eqn:f-def}).
\end{lem}
\begin{pf}
If a PSNE effort profile $(\lambda_i^*, \lambda_{-i}^*)$ exists in the given game, then it must satisfy, $u_{i,1}(\lambda_i^*, \lambda_{-i}^*) \geq u_{i,1}(\lambda_i, \lambda_{-i}^*), \forall \lambda_i \in S_i=[0,\infty), \forall i \in N$.
This implies, $\lambda_i^* = \underset{\lambda_i \in S_i=[0,\infty)}{\argmax} u_{i,1}(\lambda_i, \lambda_{-i}^*), \forall \ i \in N$.

Thus in order to find the Nash equilibrium we have to solve the following optimization problem for each $i \in N$.

\vspace{-.1in}
\begin{equation} \label{optprob}
\begin{array}{ccc}
 \left.
 \begin{array}{ll}
  \max_{\lambda_i} & u_{i,1}(\lambda_i, \lambda_{-i}^*) \\
  \text{s.t.} & \lambda_i \geq 0,
 \end{array} 
 \right \}
& \Rightarrow
& 
\begin{array}{ll}
  \min_{\lambda_i} & -u_{i,1}(\lambda_i, \lambda_{-i}^*) \\
  \text{s.t.} & -\lambda_i \leq 0
 \end{array}
\end{array}
\end{equation}
\vspace{-.15in}

\noindent Due to concavity of $u_{i,1}$, this is a convex optimization problem with linear constraints, which can be solved using KKT theorem. At the minimizer $\lambda_i^*$ of problem (\ref{optprob}), $\exists \ \mu \in \mathbb{R}$ such that, 
(i) $\mu \geq 0$, 
(ii) $-\frac{\partial}{\partial\lambda_i}u_{i,1}(\lambda_i^*, \lambda_{-i}^*) -\mu = 0$, 
(iii) $-\mu\lambda_i^* = 0$,
(iv) $-\lambda_i^* \leq 0$. \\
Case 1: $\mu>0 \Rightarrow \lambda_i^* = 0$ and in this case $\frac{\partial}{\partial\lambda_i}u_{i,1} = -\mu \leq 0$.\\
Case 2: $\mu=0 \Rightarrow \frac{\partial}{\partial\lambda_i}u_{i,1}(\lambda_i^*, \lambda_{-i}^*) = 0$ and in this case $\lambda_i^*\geq 0$. This leads us to (from Eq. (\ref{du1bydlami})),
\begin{align}
\left(\sum_{j \in N} \lambda_j^*\right)^2 =  \frac{\lambda \gamma R}{C}\left(\sum_{j \neq i} \lambda_j^* - \sum_{j \in T_i \setminus \{i\}} \lambda_j^* \delta_{ij} \right) 
\label{lami_untrunc}
\end{align}
For a given tree and its equilibrium profile $\boldsymbol \lambda^*$, let $\sum_{j \in N} \lambda_j^* = x$ (variable substitution), then manipulation of (\ref{lami_untrunc}) leads to,
\begin{align}
\lambda_i^* + \sum_{j \in T_i \setminus \{i\}} \lambda_j^* \delta_{ij} = x - \frac{x^2C}{\lambda \gamma R}, \ \forall \ i \ \in N
\label{effort_eq}
\end{align}
We do another variable substitution to denote the RHS of Eqn (\ref{effort_eq}) by $y$ ($\geq 0$ since LHS is 
$\geq 0)$. That is,
\begin{equation}
y = x - \frac{x^2C}{\lambda \gamma R}.
\label{k2}
\end{equation}
\begin{lem}
  $\lambda_i^* = yf(T_i,\Delta), \ \forall i \in N$.
\end{lem}
\begin{pf}
We prove this claim via induction on the levels of $T$. Let the depth of $T$ be $D$.

From Eqn (\ref{effort_eq}), $\lambda_i^* + \sum_{j \in T_i \setminus \{i\}}\delta_{ij}  \lambda_j^* = y, \forall \ i\in N$. From Cases 1 and 2 above, 
\begin{align}
\lambda_i^* = \textnormal{max}\left(0,y-\sum_{j \in T_i \setminus \{i\}}\delta_{ij}  \lambda_j^*\right) \ \forall \ i\in N
\label{effort_eq3}
\end{align}

\textbf{\emph{Step 1}}: For an arbitrary node $j$ at level $D$, from (\ref{effort_eq3}), $\lambda_j^* = y$. Hence, the proposition is true as $f(T_j,\Delta) = 1$ for a leaf.

Now, select an arbitrary node $i$ (which is not a leaf) at level $D-1$. From (\ref{effort_eq3}) we get, $\lambda_i^* = \max (0,y-\sum_{j \in T_i \setminus \{i\}}\delta_{ij} y) = y \max (0,1-\sum_{j \in T_i \setminus \{i\}} \delta_{ij} 1) = yf(T_i,\Delta)$.

\textbf{\emph{Step 2}}: Let $\lambda_j^* = yf(T_j,\Delta)$ be true for all nodes $j$ upto level $D-l$. Consider an arbitrary node $i$ at level $D-l-1$. From (\ref{effort_eq3}) and (\ref{eqn:f-def}),

\vspace{-.12in}
\small
\begin{align}
\lambda_i^* = \textnormal{max}\left(0,y-\sum_{j \in T_i \setminus \{i\}} y \cdot f(T_j,\Delta)\delta_{ij}\right) = yf(T_i,\Delta)
\end{align}
\normalsize
which concludes the induction.
\end{pf}
\newline
To find an expression for PSNE we now evaluate $y$. The sum of efforts of all the players is defined as $x$. Hence, $x = y\sum_{j \in N} f(T_j,\Delta)$. Substituting for $y$ from (\ref{k2}) in this expression and solving for $x$ yields,
\begin{align}
x = \sum_{j \in N} \lambda_j^* = \frac{\lambda \gamma R}{C}\left( \frac{\sum_{j \in N} f(T_j,\Delta)-1}{\sum_{j \in N} f(T_j,\Delta)} \right)
\label{k1}
\end{align}
Using (\ref{k2}) we get,
\small
\begin{align}
y = \frac{\lambda \gamma R}{C}\left( \frac{\sum_{j \in N} f(T_j,\Delta)-1}{\left(\sum_{j \in N} f(T_j,\Delta)\right)^2} \right)
\label{k2-expr}
\end{align}
\normalsize
Combining (\ref{k2-expr}) and the claim above, the PSNE is given by,
\small
\begin{align}
\lambda_i^* = \frac{\lambda \gamma R}{C}\left( \frac{\sum_{j \in N} f(T_j,\Delta) - 1}{\left(\sum_{j \in N} f(T_j,\Delta)\right)^2} \right) f(T_i,\Delta), \forall \ i \in N
\nonumber
\end{align}
\normalsize
KKT equations led to a unique solution of the optimization problem, hence PSNE is unique.
\end{pf}

\begin{thm}[PSNE in $R_4$]
\label{theorem:R4_lies_in_Z2_Z4}
 If $\mathbf{v} \in R_4$, then there exists a unique PSNE effort profile $\boldsymbol\lambda^*$ given by Eqn~(\ref{eqn:R4-lambda}), which lies in $Z_4$.
\end{thm}

\begin{pf}
From Corollary \ref{corollary:util_in_R4} when $\mathbf{v} \in R_4$, $u_i(\boldsymbol \lambda^*) = u_{i,1}(\boldsymbol \lambda^*)$. From Lemma \ref{lemma:R4-exact-expr} with this utility function, the unique PSNE effort profile is given by Eqn (\ref{eqn:R4-lambda}). We use this expression to compute the sum $\sum_{j \in N} \lambda_j^*$, and substitute the value of $\gamma R/C$ from (\ref{eqn:R4}) to get $\sum_{j \in N} \lambda_j^* > \lambda \Rightarrow \boldsymbol\lambda^* \in Z_4$.
\end{pf}

In $Z_4$, the sum of the equilibrium efforts is more than $\lambda$. Hence, it shows that in order to meet the goal of efficiently serving the incoming tasks, the reward to cost ratio has to be above a threshold given by (\ref{eqn:R4}). We also see that when the parameters are in $R_4$, the equilibrium effort of agent $i$ is proportional to $f(T_i,\Delta)$.

\textbf{Now we discuss region $R_3$.} In region $R_3$, the reward to cost ratio is a little larger than $R_2$. However, it is still not enough to guarantee the sum equilibrium effort levels to cross $\lambda$ (which was the case in $R_4$). In particular, we show that it is necessary and sufficient for the equilibrium effort profile to live in $Z_3$. To show that we need $Z_3$ to be nonempty.

\begin{lem} \label{lemma:Z_3-nonempty}
 If $\mathbf{v} \in R_3$, $Z_3$ is nonempty.
\end{lem}
\begin{pf}
The proof is constructive. Let us consider the following summation: $S = \lambda \left(1-\frac{C}{\gamma R}\right)\sum_{j \in N}f(T_j,\Delta)$. Substituting $\frac{C}{\gamma R} \geq \frac{\sum_{j \in N}f(T_j,\Delta) - 1}{\sum_{j \in N}f(T_j,\Delta)}$ (using (\ref{eqn:R3})), we get $S \leq \lambda$. Now, let us construct a $\boldsymbol \lambda$, such that $\lambda_i = \lambda \left(1-\frac{C}{\gamma R}\right) f(T_i,\Delta) + \beta_i(\lambda - S), \forall \ i \in N$, where, $\beta_i \geq 0,\ \forall i \in N$ and $\sum_{i=1}^n\beta_i = 1$. Hence, by construction, $\sum_{i \in N} \lambda_i = \lambda \Rightarrow \boldsymbol \lambda \in Z_2$. We see that, 

{\scriptsize
\begin{align*}
  \lambda_i + \sum_{j \in T_i \setminus \{i\}} \delta_{ij} \lambda_j &= \underbrace{\left(f(T_i,\Delta) + \sum_{j \in T_i \setminus \{i\}} \delta_{ij} f(T_j,\Delta) \right)}_{\geq 1, \text{ using definition of $f$ from Eq.~(\ref{eqn:f-def})}} \cdot \lambda \left(1-\frac{C}{\gamma R}\right) \\
 \qquad &+ \underbrace{\left( \beta_i + \sum_{j \in T_i \setminus \{i\}} \delta_{ij} \beta_j \right) (\lambda - S)}_{\geq 0} \\
 & \geq \lambda \left(1-\frac{C}{\gamma R}\right)
\end{align*}
}

\noindent So, $\boldsymbol \lambda \in Z_3$, hence $Z_3$ is nonempty.
\end{pf}

\begin{thm}[Necessary and Sufficient Condition for PSNE in $R_3$]
\label{thm:R3-in-Z2}
If $\mathbf{v} \in R_3$, an effort profile $\boldsymbol\lambda^*$ is a PSNE, if and only if $\boldsymbol\lambda^* \in Z_3$.
\end{thm}
\begin{pf} ($\Leftarrow$): We show that if $\boldsymbol\lambda^* \in Z_3$, it is a PSNE. From Lemma~\ref{lemma:Z_3-nonempty}, $Z_3 \neq \phi$, so we can pick a $\boldsymbol\lambda^* \in Z_3$. Since, $\boldsymbol\lambda^* \in Z_3$ implies $\boldsymbol\lambda^* \in Z_2$, the utility undergoes a transition at this point, $u_i(\lambda_i^*-,\lambda_{-i}^*) = u_{i,2}(\lambda_i^*-,\lambda_{-i}^*)$, and $u_i(\lambda_i^*+,\lambda_{-i}^*) = u_{i,1}(\lambda_i^*+,\lambda_{-i}^*)$. Since $u_i$ is continuous but not differentiable at $\boldsymbol\lambda^*$, we look at the left and right derivatives at this point.

{\small
\begin{align*}
 \frac{\partial}{\partial \lambda^*_i}u_{i}(\lambda^*_i-, \lambda^*_{-i}) &= \frac{\partial}{\partial \lambda^*_i}u_{i,2}(\lambda^*_i-, \lambda^*_{-i}) = \gamma R - C > 0, \\
  \frac{\partial}{\partial \lambda^*_i}u_{i}(\lambda^*_i+, \lambda^*_{-i}) &= \frac{\partial}{\partial \lambda^*_i}u_{i,1}(\lambda^*_i+, \lambda^*_{-i}) \\
 &= \frac{\gamma R \left(\lambda - \lambda_i^* - \sum_{j \in T_i \setminus \{i\}} \delta_{ij} \lambda_j^* \right)}{\lambda} - C \leq 0
\end{align*}
}

The first inequality comes because $\mathbf{v} \in R_3$. Since $\boldsymbol\lambda^* \in Z_3 \subseteq Z_2$, the equality in the third line comes by replacing $\sum_{j \in N} \lambda_j^* = \lambda$ in Eq.~(\ref{du1bydlami}), and the inequality in the third line is obtained by reorganizing the expression in Eqn (\ref{eqn:def-Z_3}). Hence for each agent $i$, the utility is maximized at $\boldsymbol\lambda^*$, when other players are playing the equilibrium strategy. So, $\boldsymbol\lambda^*$ is a PSNE.

($\Rightarrow$): We are given that $\mathbf{v} \in R_3$, and $\boldsymbol\lambda^*$ is a PSNE. We first show that $\boldsymbol\lambda^*$ should necessarily be in $Z_2$. We show this via contradiction. Let us consider the two following cases.

\textbf{\emph{Case 1}}: Let $\boldsymbol \lambda^*$ be the PSNE if it exists such that $\boldsymbol\lambda^* \in Z_1$, that is, $\sum_{j \in N} \lambda_j^* < \lambda$ then, $u_i(\boldsymbol\lambda^*) = u_{i,2}(\boldsymbol\lambda^*), \forall \ i \in N$. But, $\frac{\partial}{\partial\lambda_i}u_{i,2}(\boldsymbol \lambda) = \gamma \cdot R - C > 0, \forall \ i \in N$ (from (\ref{eqn:R3})). Hence, utility of $i$ is increasing in $\lambda_i$ at $\boldsymbol \lambda^*$. So, $i$ is better off by increasing his effort from $\lambda_i$, which is a contradiction to the fact that $\lambda^*$ is a PSNE.

\textbf{\emph{Case 2}}: Let $\boldsymbol \lambda^*$ be the PSNE if it exists such that $\boldsymbol\lambda^* \in Z_3$, that is, $\sum_{j \in N} \lambda_j^* > \lambda$ then, $u_i(\boldsymbol\lambda^*) = u_{i,1}(\boldsymbol\lambda^*), \forall \ i \in N$. Thus Eqn (\ref{k1}) is valid. Using (\ref{eqn:R3}) in (\ref{k1}) we get $\sum_{j \in N} \lambda_j^* \leq \lambda$ which is a contradiction.
Hence, if the PSNE effort profile $\boldsymbol\lambda^*$ exists, it must lie in $Z_2$.

Now, since $\boldsymbol\lambda^* \in Z_2$, the utility will have a transition at this point. We know that $\boldsymbol\lambda^*$ is a PSNE, hence the utility must be maximized at this point for all agents $i \in N$, given that the other agents stick to the equilibrium strategies. Hence, it must hold that,

\vspace{-.1in}
{\small
\begin{align}
 \frac{\partial}{\partial\lambda_i}u_i(\lambda_i^*-, \lambda_{-i}^*) &= \frac{\partial}{\partial\lambda_i}u_{i,2}(\lambda_i^*, \lambda_{-i}^*) \geq 0, \mbox{ and } \label{cond-1} \\
 \frac{\partial}{\partial\lambda_i}u_i(\lambda_i^*+, \lambda_{-i}^*) &= 
 \frac{\partial}{\partial\lambda_i}u_{i,1}(\lambda_i^*, \lambda_{-i}^*)
 \leq 0. \label{cond-2}
\end{align}
}
Condition (\ref{cond-1}) yields, $\gamma \cdot R - C \geq 0$, which is satisfied since $\mathbf{v} \in R_3$. Condition (\ref{cond-2}) yields, $\frac{\gamma R \left(\lambda - \lambda_i^* - \sum_{j \in T_i \setminus \{i\}} \delta_{ij} \lambda_j^* \right)}{\lambda} - C \geq 0$, since we have shown that $\boldsymbol\lambda^* \in Z_2$. Reorganizing this inequality, we get,
$\lambda_i^* + \sum_{j \in T_i \setminus \{i\}} \delta_{ij} \lambda_j^* \geq \lambda \left( 1 - \frac{C}{\gamma R} \right)$. Hence, $\boldsymbol\lambda^* \in Z_3$.
\end{pf}

\begin{figure*}
\centering
\subfloat[Referral Tree. \label{fig:tree}]{
\includegraphics[width=0.33\textwidth]{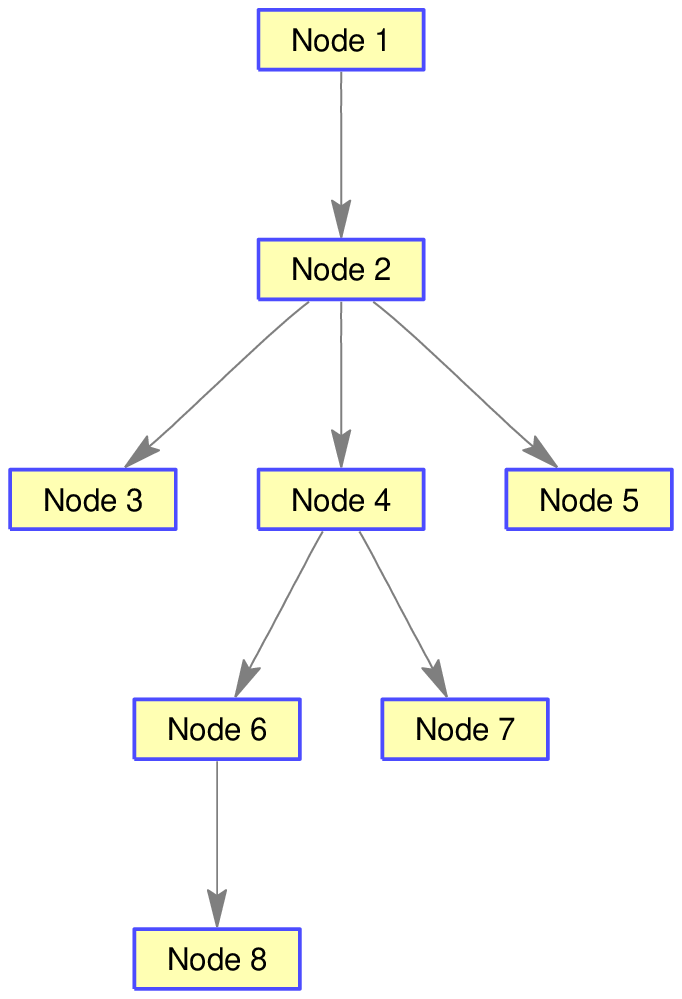} }
\subfloat[Effort and rewards, $a = 2$. \label{fig:effortutility-a2}]{
\includegraphics[width=0.33\textwidth]{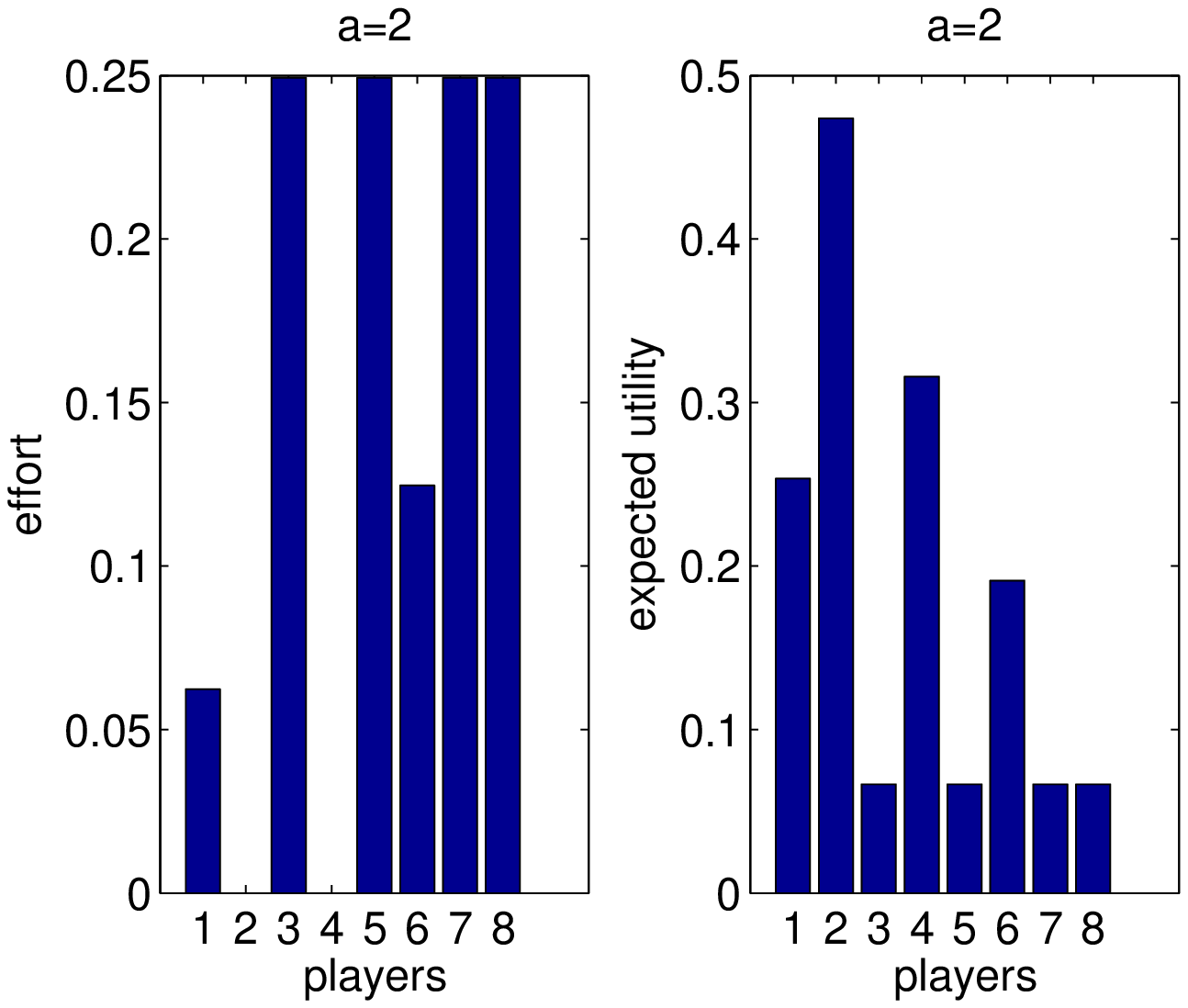} }
\subfloat[Effort and rewards, $a = 3$. \label{fig:effortutility-a3}]{
\includegraphics[width=0.33\textwidth]{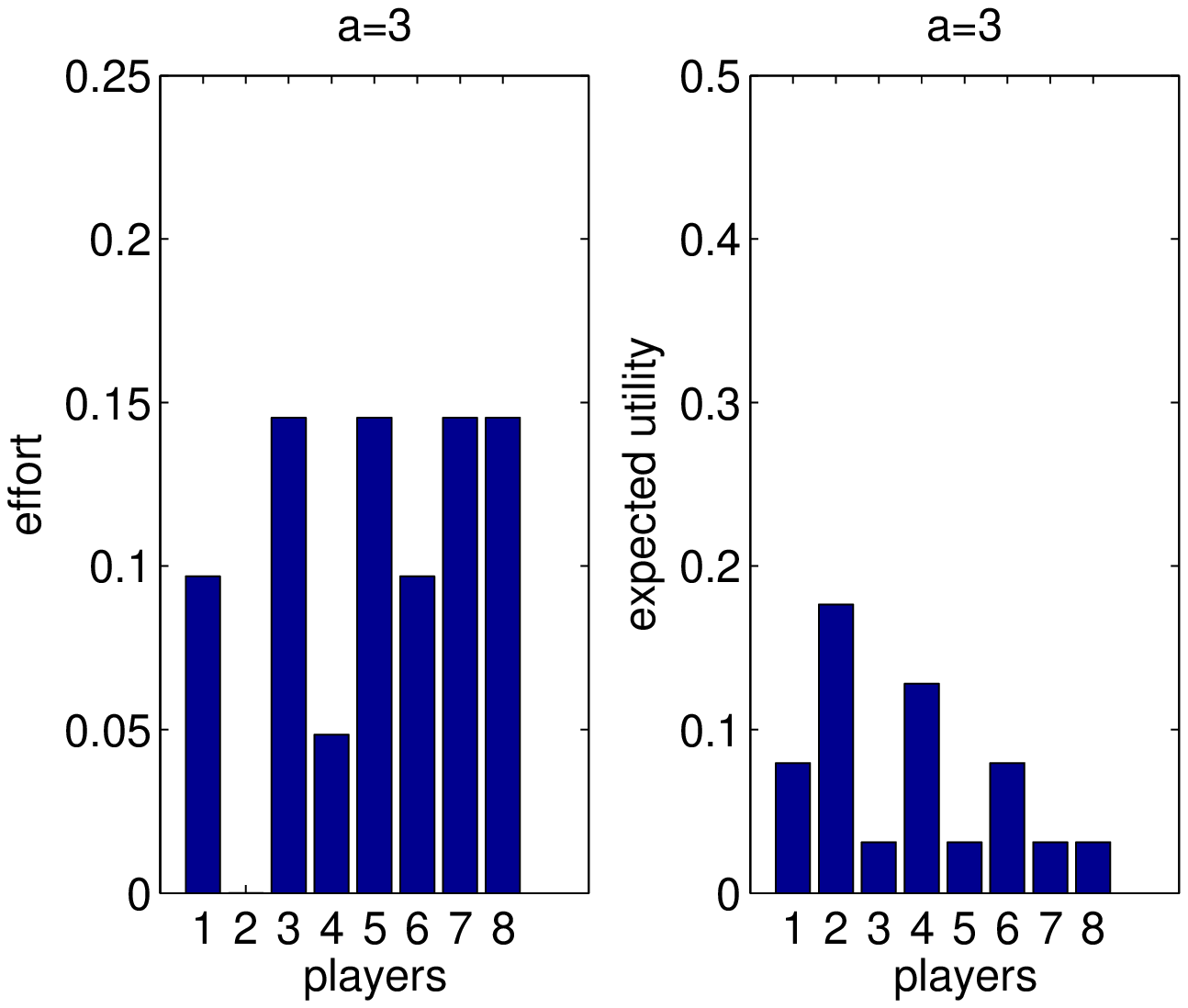} }
\caption{\small Referral tree and PSNE effort/reward profiles for the referral tree shown in Fig.\ref{fig:tree}. $\lambda = 0.2$, $R=15$, $C=1$.}
\label{fig:tree_effortutility}
\end{figure*}

\textbf{To summarize this section}, in the model considered, pure strategy Nash equilibria always exist.  The regions correspond to different reward to cost ratio, and the results predict the efforts expected from rational agents when the ratio lies in one of these regions. It gives a measure of how large $\gamma R/C$ needs to be in order to efficiently serve the incoming task process (to operate in $Z_4$). The results suggest that nodes having network advantages due to recruiting other nodes free ride on others.  Thus, a geometric incentive mechanism may disincentivize nodes that have large subtrees from putting efforts.

\section{Numerical Results} \label{sec:numerical}

We numerically compute the PSNE  (using Theorem \ref{theorem:R4_lies_in_Z2_Z4}) to illustrate the effect of the referral tree on individual efforts in region $R_4$ --- the most interesting region where all incoming tasks are served. We consider a referral tree shown in Fig. \ref{fig:tree}. To illustrate the analytical results we enforce the structure in Eqn (\ref{geo_reward}) on the reward sharing matrix $\Delta = [ \delta_{ij}]$. This is motivated by the geometric incentive mechanism used by the winning team in the DARPA Network challenge \cite{Pickard2011}.
\begin{align}
\delta_{ij} = \left\{
\begin{array}{ll}
\left( \frac{1}{a}\right )^{\text{dist}_T(i,j) + 1   }  & j \in T_i,\\ 
0 & \text{otherwise.}
\end{array}
\right. \label{geo_reward}
\end{align}
Here, $a > 1 $ and $\gamma=\delta_{ii}=1/a$. The PSNE effort profiles and expected utility are shown in Fig. \ref{fig:effortutility-a2} and Fig. \ref{fig:effortutility-a3}. As seen in both the figures, node $2$ who has three children does not put any effort due to passive rewards.  In the $a=2$ case, node $4$ does not put any effort, but when the passive reward is reduced by increasing  $a$ to $3$, node $4$ puts in some effort. Although the sum effort is higher when $a=2$, the variability in the effort among the nodes is also high compared to $a=3$.

\begin{figure}
\begin{center}
\includegraphics[width=0.35\textwidth]{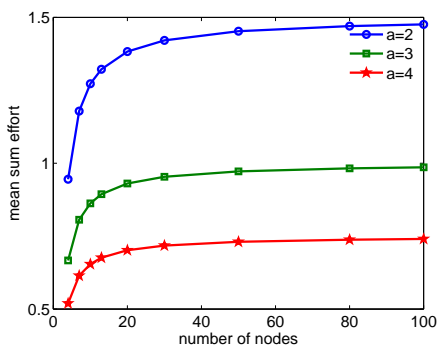}
\end{center}
\caption{\small Mean sum effort vs. number of nodes for varying $a$. Trees are generated randomly. Results obtained by averaging over 500 samples. $\lambda = 0.2$, $R=15$, $C=1$.}
\label{fig:sumeffort}
\end{figure}

Fig. \ref{fig:sumeffort} illustrates the effect of $a$ on sum effort. Trees were generated randomly. To generate a tree of size $N$ we carry out the following process recursively. Given a tree of size $i$, the $(i+1)^{th}$ node chooses a parent at random from $i$ nodes to attach itself. We discover that the sum effort saturates as number of nodes in the tree increases. This suggests that recruiting large number of individuals using geometric incentives may not yield proportional increase in productivity.

\section{Conclusions} \label{sec:concl}

In this paper we performed a game theoretic analysis on the efforts put in by individuals recruited using a referral tree, which is a popular crowd sensing mechanism for recruiting individuals. We propose a queuing model with Poisson task arrivals where agents compete to finish the task. Agents receive not only direct rewards for finishing a task, but also indirect rewards if any agent in their subtree finishes a task. We provide a complete analysis and a closed form solution for all possible system parameters. We compute the PSNE effort profile for the complete parameter space and show that the PSNE always exists. In some regions of the parameter space it is unique, while in others it is not.  Our results uncover free riding behavior among nodes who obtain large passive rewards from their subtrees. This has implications on crowd sourced tasks such as the DARPA Network challenge. In particular, usage of geometric incentive mechanisms to recruit large number of individuals may not result in proportionate effort due to free riding.

\small
\vspace{-.07in}
\section*{\small{Acknowledgement}}
\vspace{-.07in}
This work was a part of the course project for the ``Game Theory'' course which the authors took during January--April 2012 session at IISc. Authors thank the instructor Prof. Y Narahari and the TA Swaprava Nath for introducing them to DARPA Network Challenge and for their comments on improving the presentation of this paper.
\normalsize

\footnotesize
\bibliographystyle{IEEEtran}
\bibliography{comsnets_snw_Bib}

\end{document}